\documentclass[sigconf]{acmart}
\renewcommand\footnotetextcopyrightpermission[1]{}
\copyrightyear{2026}
\acmYear{2026}
\acmDOI{}
\acmConference[ACM MM '26]{Proceedings of the 34th ACM International Conference on Multimedia}{November 2026}{Brazil}
\acmISBN{}

\usepackage{booktabs}
\usepackage{multirow}
\usepackage{amsmath}
\usepackage{graphicx}
\usepackage{subcaption}
\usepackage{xcolor}
\usepackage{enumitem}
\usepackage{array}
\usepackage{microtype}

\title{FIRMED: A Peak-Centered Multimodal Dataset with Fine-Grained Annotation for Emotion Recognition}

\author{Hao Tang}
\authornote{Both authors contributed equally to this research.}
\affiliation{%
  \institution{Northwestern Polytechnical University}
  \city{Xi'an}
  \country{China}}
\email{haotang@mail.nwpu.edu.cn}

\author{Songyun Xie}
\authornote{Corresponding authors.}
\affiliation{%
  \institution{Northwestern Polytechnical University}
  \city{Xi'an}
  \country{China}}
\email{syxie@nwpu.edu.cn}

\author{Xinzhou Xie}
\authornotemark[2] 
\affiliation{%
  \institution{Northwestern Polytechnical University}
  \city{Xi'an}
  \country{China}}
\email{xinzhxie@nwpu.edu.cn}

\author{Can Liao}
\authornotemark[1] 
\affiliation{%
  \institution{Northwestern Polytechnical University}
  \city{Xi'an}
  \country{China}}
\email{liao_can@mail.nwpu.edu.cn}

\author{Bohan Li}
\affiliation{%
  \institution{Northwestern Polytechnical University}
  \city{Xi'an}
  \country{China}}
\email{bhli@mail.nwpu.edu.cn}

\author{Zhongyu Tian}
\affiliation{%
  \institution{Northwestern Polytechnical University}
  \city{Xi'an}
  \country{China}}
\email{tzy2157@mail.nwpu.edu.cn}

\author{Dalu Zheng}
\affiliation{%
  \institution{Northwestern Polytechnical University}
  \city{Xi'an}
  \country{China}}
\email{zhengdl@mail.nwpu.edu.cn}
\begin{document}

\begin{abstract}
Traditional video-induced physiological datasets usually rely on whole-trial labels, which introduce temporal label noise in dynamic emotion recognition. We present FIRMED, a peak-centered multimodal dataset based on an immediate-recall annotation paradigm, with synchronized EEG, ECG, GSR, PPG, and facial recordings from 35 participants. FIRMED provides event-centered timestamps, emotion labels, and intensity annotations, and its annotation quality is supported by subjective and physiological validation. Benchmark experiments show that FIRMED consistently outperforms whole-trial labeling, yielding an average gain of 3.8 percentage points across eight EEG-based classifiers, with further improvements under multimodal fusion. FIRMED provides a practical benchmark for temporally localized supervision in multimodal affective computing.
\end{abstract}


\maketitle

\section{Introduction}

Emotion recognition plays a pivotal role in affective computing and multimodal multimedia analysis, driving advancements in human-computer interaction, affect-aware recommendation, and intelligent content understanding~\cite{soleymani2011multimodal, pillalamarri2025review}. However, recognizing emotions from dynamic video stimuli presents a unique challenge: human affective responses are rarely static; rather, they evolve continuously and are often triggered by specific, transient salient events. While recent studies increasingly integrate behavioral cues (e.g., facial expressions) with physiological signals (e.g., EEG, ECG, GSR, PPG) to capture these dynamics comprehensively, the quality of the underlying supervision remains insufficiently addressed~\cite{zhang2022eeg}. Most existing datasets assign a single global label to an entire video trial. This coarse, whole-trial labeling strategy inherently introduces temporal label noise by conflating emotionally salient moments with neutral or irrelevant periods, thereby degrading the learning efficacy of downstream recognition models.

To mitigate this issue, recent efforts have explored continuous or fine-grained annotation strategies~\cite{jiang2024seed,wang2023mgeed}. Yet, precisely aligning subjective emotional experiences with rapidly fluctuating multimodal signals remains profoundly difficult in dynamic video settings. Retrospective or delayed annotations are particularly susceptible to temporal uncertainty and memory bias, as participants struggle to faithfully reconstruct their continuously varying emotions post-exposure. This limitation is consistent with the \textit{Peak-End Rule} from cognitive psychology~\cite{kahneman1993more}, which posits that retrospective evaluations are disproportionately dominated by the most intense moments (peaks) of an experience, rather than its full temporal trajectory. Consequently, both conventional whole-trial labeling and continuous retrospective annotation are prone to injecting label noise into video-induced emotion recognition systems.

Motivated by these insights, we introduce \textbf{FIRMED}, a \textbf{F}ine-grained \textbf{I}mmediate \textbf{R}ecall-based \textbf{M}ultimodal \textbf{E}motion \textbf{D}ataset. This dataset utilizes a peak-centered annotation paradigm wherein participants report their self-perceived emotional peaks immediately following each stimulus. Instead of relying on global trial-level labels, this paradigm captures event-centered timestamps alongside emotion categories and intensity levels, delivering highly localized supervision. FIRMED comprises ecologically valid short-video stimuli paired with synchronized EEG, ECG, GSR, PPG, and facial recordings from 35 participants. Additionally, recognizing that emotional reactivity is inherently modulated by individual differences, we concurrently collected Big Five personality profiles to account for subjective variations. Through extensive physiological analyses and benchmark experiments, we show that this peak-centered annotation strategy consistently improves downstream multimodal emotion recognition.

The main contributions of this work are summarized as follows:
\begin{itemize}
    \item \textbf{A Peak-Centered Emotion Annotation Paradigm:} We propose an immediate-recall, peak-centered annotation paradigm that provides temporally localized supervision and mitigates the limitations of conventional whole-trial labeling.
    \item \textbf{The FIRMED Dataset:} We construct \textbf{FIRMED}, a multimodal dataset of ecologically valid short-video stimuli with synchronized EEG, ECG, GSR, PPG, facial recordings, Big Five personality scores, and precise event-centered annotations.
    \item \textbf{Comprehensive Validation and Benchmarking:} We perform subjective, physiological, and computational evaluations to validate the proposed annotation strategy, and show that peak-centered supervision consistently outperforms whole-trial labeling.
\end{itemize}

\section{Related Work}

\subsection{Evolution of Datasets and Label Noise}
Early influential datasets, including DEAP \cite{koelstra2011deap}, MAHNOB-HCI \cite{soleymani2011multimodal}, and the SEED series \cite{zheng2015investigating,zheng2018emotionmeter,liu2021comparing}, established the methodological foundation for physiological emotion analysis. However, as summarized in Table \ref{tab:emotion_datasets_comparison}, these datasets universally adopted whole-trial labeling, assigning a single emotion label to a multi-minute stimulus segment. While this strategy enabled large-scale benchmark development, it fails to capture the temporally evolving nature of emotions. Emotional reactions are typically triggered by specific salient events rather than being uniformly sustained \cite{hakim2013computational}. Consequently, coarse whole-trial labeling introduces temporal label noise by forcing non-salient or neutral physiological segments to act as positive samples \cite{jiang2024remonet}.

To improve temporal granularity, recent datasets like MGEED \cite{wang2023mgeed} and SEED-VII \cite{jiang2024seed} introduced fine-grained annotation strategies (e.g., frame-level or 4-s window delayed ratings). While representing progress, these retrospective paradigms suffer from temporal uncertainty and memory bias. As suggested by the psychological \textit{Peak-End Rule} \cite{kahneman1993more, fredrickson1993duration}, retrospective evaluations are disproportionately influenced by the most intense or final moments, rather than a faithful temporal reconstruction. Thus, overcoming these memory biases to achieve precise emotion annotation remains the foremost challenge and consideration in dataset construction.

\begin{table*}[!htbp]
\centering
\caption{Evolution of Multimodal Emotion Datasets: Progression of Labeling Granularity}
\label{tab:emotion_datasets_comparison}
\renewcommand{\arraystretch}{1.3} 
\setlength{\tabcolsep}{5pt}
\footnotesize
\begin{tabular}{>{\centering\arraybackslash}p{0.8cm} >{\raggedright\arraybackslash}p{2.0cm} >{\raggedright\arraybackslash}p{2.8cm} >{\raggedright\arraybackslash}p{3.2cm} >{\raggedright\arraybackslash}p{3.0cm} >{\raggedright\arraybackslash}p{3.8cm}}
\toprule
\textbf{Year} & \textbf{Dataset} & \textbf{Experimental Setup} & \textbf{Signals} & \textbf{Emotion Model} & \textbf{Labeling Strategy} \\
\midrule
\multicolumn{6}{c}{\textbf{Coarse-Grained Annotation Paradigm}} \\
\midrule
2012 & \textbf{DEAP} \cite{koelstra2011deap} & $N$=32; 40 clips (60s) & EEG (32), EOG, EMG, GSR & VAD & \textbf{Whole-trial} (Single label) \\
2013 & \textbf{SEED} \cite{zheng2015investigating} & $N$=15; 45 clips ($\sim$4m) & EEG (62), EOG & Pos-Neg-Neutral & \textbf{Whole-trial} (Single label) \\
2017 & \textbf{DREAMER} \cite{katsigiannis2017dreamer} & $N$=23; 18 clips (60s) & EEG (14), ECG & VAD & \textbf{Whole-trial} (Single label) \\
2018 & \textbf{AMIGOS} \cite{miranda2018amigos} & $N$=40; 20 clips & EEG (14), ECG, GSR & VAD & \textbf{Whole-trial} (Single label) \\
2019 & \textbf{MPED} \cite{song2019mped} & $N$=23; 28 clips & EEG (62), ECG, GSR & 7 Discrete & \textbf{Whole-trial} (Single label) \\
2021 & \textbf{SEED-V} \cite{liu2021comparing} & $N$=20; 45 clips & EEG (62), EOG & 5 Discrete + Neutral & \textbf{Whole-trial} (Single label) \\
\midrule
\multicolumn{6}{c}{\textbf{Fine-Grained Annotation Paradigm}} \\
\midrule
2023 & \textbf{MGEED} \cite{wang2023mgeed} & $N$=17; 9 clips ($\sim$4m) & EEG (14), ECG, GSR & 6 Discrete + VAD & \textbf{Frame-level} (Delayed recall) \\
2024 & \textbf{SEED-VII} \cite{jiang2024seed} & $N$=20; 80 clips (2-5m) & EEG (62), EOG & 6 Discrete + Neutral & \textbf{4-s windows} (Delayed recall) \\
\bottomrule
\end{tabular}
\vspace{0.1cm}
\begin{minipage}{\textwidth}
\textit{Note:} VAD = Valence-Arousal-Dominance. 
\end{minipage}
\end{table*}

\subsection{Physiological Emotion Recognition}
Alongside dataset evolution, emotion classification algorithms have advanced significantly. In single-modality EEG analysis, the field has transitioned from traditional machine learning \cite{wang2011eeg} to end-to-end deep learning. Architectures like EEGNet \cite{lawhern2018eegnet} extract spatio-temporal features, while Graph Convolutional Networks (GCNs) effectively model brain functional connectivity as dynamic graphs \cite{song2018eeg, tang2024multi}.

However, emotions are complex, multi-system phenomena involving both the Central Nervous System (CNS) \cite{shen2025ua} and the Autonomic Nervous System (ANS) \cite{shen2025physiological, kreibig2010autonomic}. Consequently, the dominant trend is multimodal fusion, combining EEG with peripheral signals (GSR, ECG) \cite{goshvarpour2017accurate, wang2023novel} and eye movements \cite{lu2015combining, gong2024ciabl} to achieve robust classification. Recent works have introduced fusion architectures to handle multimodal discrepancies. For instance, UAGCFNet \cite{li2025uncertainty} employs an Uncertainty-Aware GCN alongside Transitive Contrastive Fusion to mitigate inter-modal uncertainty and unimodal bias. As these architectures become increasingly complex, providing them with temporally precise, noise-free supervision labels becomes critical for realizing their full predictive potential.
\section{FIRMED Dataset}

\subsection{Stimuli and Participants}
To maximize ecological validity, FIRMED utilizes short videos sourced from social media platforms (e.g., Douyin, YouTube). These stimuli inherently possess higher information density and elicit more rapid emotional fluctuations than traditional movie clips. A panel of trained raters rigorously screened candidate videos, retaining only those achieving unanimous agreement on eliciting one of six discrete emotions (Happiness, Sadness, Anger, Fear, Disgust, and Surprise; 15 videos each, 90 total). 

Thirty-five healthy participants (20 males, 15 females; mean age $23.9 \pm 1.7$ years) were recruited under an ethically approved protocol with written informed consent. Meanwhile, participants completed a Big Five personality assessment prior to recordings.

\subsection{Immediate-Recall Protocol}
\begin{figure*}[htbp]   
    \centering
    \includegraphics[width=1.9\columnwidth]{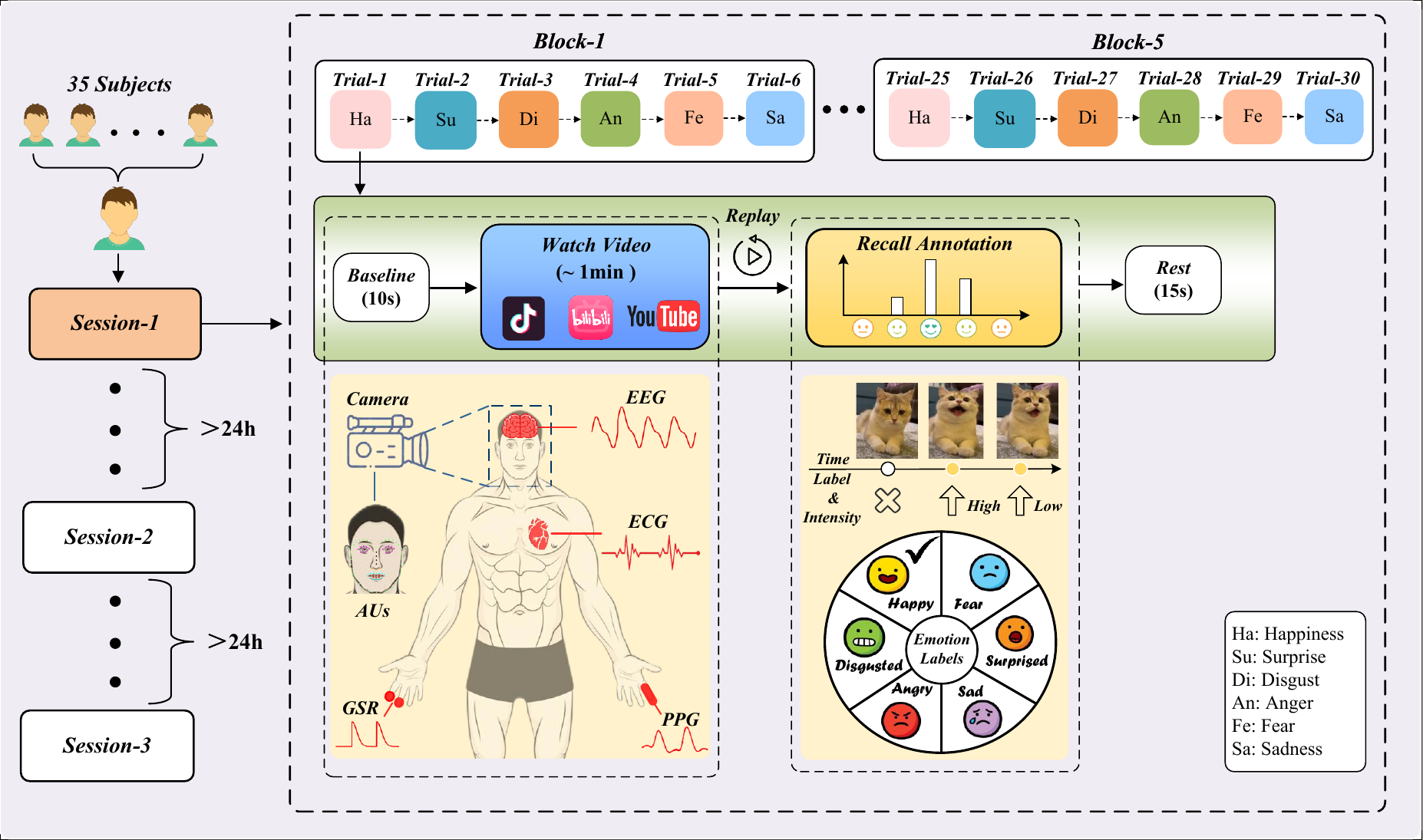}
    \caption{Experimental paradigm for FIRMED data collection, including the multi-session design, the procedure within a single trial (baseline, stimulus presentation, immediate replay with discrete annotation, and rest), the multimodal signals recorded, and the three annotation dimensions (label, intensity, and time).}
    \label{fig1}
\end{figure*}

As illustrated in Fig. \ref{fig1}, the experiment spanned three separate sessions per participant (>24-hour intervals to prevent emotional fatigue). Each session comprised 30 trials with stimuli presented in a fixed emotional sequence, repeated five times. 

% [此处保留你原来的 fig1 代码]

Each trial proceeded as follows: 

(1) \textit{Baseline}: A 10-s fixation cross for resting-state recording. 

(2) \textit{Stimulus Presentation}: A $\sim$1-minute video was viewed naturally while physiological and facial signals were continuously recorded. 

(3) \textit{Immediate Replay and Annotation}: To prevent motion artifacts from finger-mounted sensors (GSR/PPG), the experimenter operated the annotation software based on the participant's verbal cues. During the replay, the experimenter navigated to self-reported salient moments, recorded the event timestamp ($t_{event}$), and inputted the corresponding emotion label and intensity (Low/Medium/High). This paradigm allowed multiple precise timestamps to be recorded within a single trial.

Experimenter-mediated annotation was necessitated by finger-mounted GSR/PPG sensors that are highly susceptible to motion artifacts. To quantify its fidelity, we conducted two analyses. For inter-rater agreement, two independent experimenters annotated a random subset of 150 trials from audio-video recordings of participants' verbal reports: Cohen's $\kappa = 0.96$ for emotion category, with 97.7\% of paired timestamps falling within a 1-second tolerance. For operator latency---the delay between the participant's verbal cue and the experimenter's click---the mean across all 3150 trials was $387 \pm 142$~ms, well within the $\pm$2\,s tolerance of our event-centered window. Potential order effects associated with the fixed emotional sequence were further analyzed in supplementary, where no significant systematic effects were observed across baseline physiological arousal, annotation behavior, or subjective intensity ratings.

\subsection{Signal Acquisition and Preprocessing}\label{sec3.3}
Synchronized multimodal data were acquired via the Neuracle NeuroHUB platform, including 59-channel EEG (1000 Hz), 3-channel ECG (1000 Hz), 2-channel GSR (100 Hz), 1-channel PPG (50 Hz), and frontal facial videos. 

Raw signals underwent standard preprocessing. EEG data were downsampled to 250 Hz, band-pass filtered (0.5--70 Hz) with a 50 Hz notch, and subjected to ICA for artifact removal. Peripheral signals were appropriately band/low-pass filtered according to their respective physiological frequency bands to eliminate baseline drift and high-frequency noise. Frontal facial videos were processed using OpenFace 2.0~\cite{baltruvsaitis2016openface} to extract frame-level Action Unit (AU) intensities based on the Facial Action Coding System (FACS). Seventeen AUs were retained (AU01, AU02, AU04, AU05, AU06, AU07, AU09, AU10, AU12, AU14, AU15, AU17, AU20, AU23, AU25, AU26, AU45), selected for their established relevance to the six target emotions. Frames with detection confidence below 0.8 were excluded. For each event-centered window, four statistical descriptors (mean, standard deviation, maximum, and range) were computed per AU, yielding a 68-dimensional facial feature vector.

\subsection{Event-Centered Windows}
\begin{figure*}[htbp]   
    \centering
    \includegraphics[width=2\columnwidth]{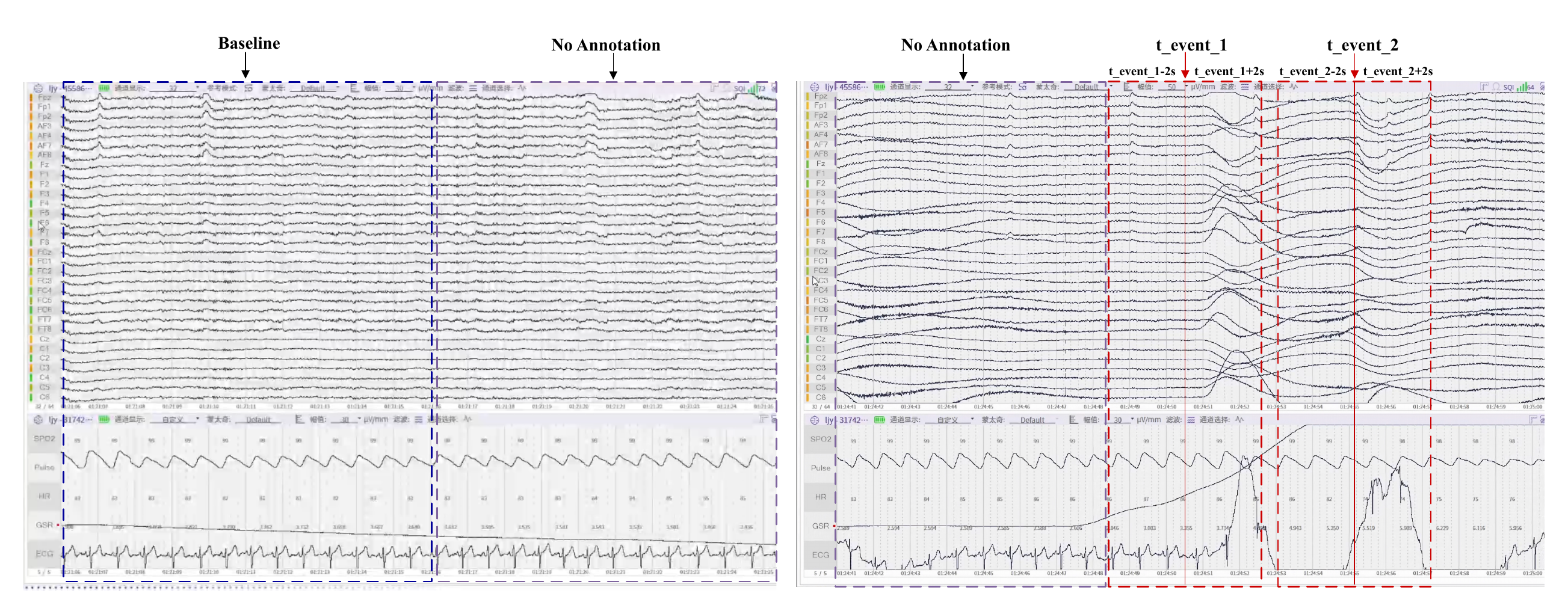}
    \caption{Raw multimodal physiological signals from a fear trial. Compared with baseline and non-annotated segments, pronounced and synchronized EEG, GSR, and ECG responses are concentrated around the self-reported peak events ($t_{event1}$, $t_{event2}$). This contrast illustrates why assigning a single trial-level label to the entire sequence may introduce temporal label noise, and motivates the use of event-centered windows for localized supervision.}
    \label{fig2}
\end{figure*}

Conventional datasets map a single label $y$ to an entire trial's physiological sequence $S$, formulated as $\mathcal{D}_{whole}=\{(x_t, y)\}_{t=1}^{T}$. This forces neutral or irrelevant segments to act as noisy positive samples. As illustrated in Fig.~\ref{fig2}, salient physiological responses are concentrated around a few self-reported peak events, whereas baseline and non-annotated segments remain comparatively uninformative. This contrast highlights why assigning a single trial-level label to the entire sequence may introduce temporal label noise.

Conversely, FIRMED extracts a tightly localized 4-s window anchored around each subjective peak timestamp:
\begin{equation}
X_{event} = [t_{event}-2\text{s},\; t_{event}+2\text{s}]
\end{equation}
This event-centered formulation ($\mathcal{D}_{event}=\{(x_{t_{event}}, y, a)\}$) provides high-confidence anchors for localized supervision. It accommodates minor subjective reporting jitter while precisely capturing the salient physiological surge, a design choice validated by our subsequent physiological analyses. Across 35 participants and 90 stimuli, FIRMED comprises a total of 14,490 annotated emotional events (detailed per-emotion and per-subject distributions are provided in the supplementary material). On average, each trial yielded 4.6 event annotations. The intensity distribution across Low, Medium, and High levels is approximately 22\%, 47\%, and 31\%, respectively.

\subsection{Objective Validation of Subjective Annotations}
While the immediate-recall paradigm minimizes memory bias, subjective self-reporting inherently carries the risk of temporal jitter or perceptual subjectivity. To establish FIRMED as a rigorous benchmark, it is crucial to demonstrate that the subjective emotional timestamps ($t_{event}$) precisely align with objective, involuntary physiological manifestations. Therefore, we conducted a cross-modal validation by analyzing both central nervous system (EEG) and autonomic nervous system (GSR) responses anchored at these self-reported peaks.

\textbf{Central Nervous System (EEG):} To investigate the neural activity patterns associated with the annotated event markers, we analyzed event-related changes in Power Spectral Density (PSD) comparing the event-centered time windows against a 10-second baseline period \cite{liu2017real}. Utilizing Welch's method, PSD was computed across five standard frequency bands: $\delta$ (1--4 Hz), $\theta$ (4--8 Hz), $\alpha$ (8--13 Hz), $\beta$ (13--30 Hz), and $\gamma$ (30--45 Hz). 

The relative PSD change was calculated as defined in Eq. \eqref{eq:eq1}. To evaluate statistical significance while rigorously addressing the multiple comparisons problem across 59 channels, we applied cluster-based permutation testing (1,000 permutations, $p<0.05$) \cite{maris2007nonparametric}. This non-parametric approach effectively controls the family-wise error rate, enhancing sensitivity to spatially coherent neural signatures. The resulting group-level patterns are visualized in Fig.~\ref{fig17}, while detailed frequency-band and regional summaries are provided in Supplementary Table~S1.

\begin{equation}
 \Delta PSD=PSD_{event}-PSD_{baseline}
 \label{eq:eq1}
\end{equation}

\begin{figure}[htbp]
	\centering
	\includegraphics[width=\columnwidth]{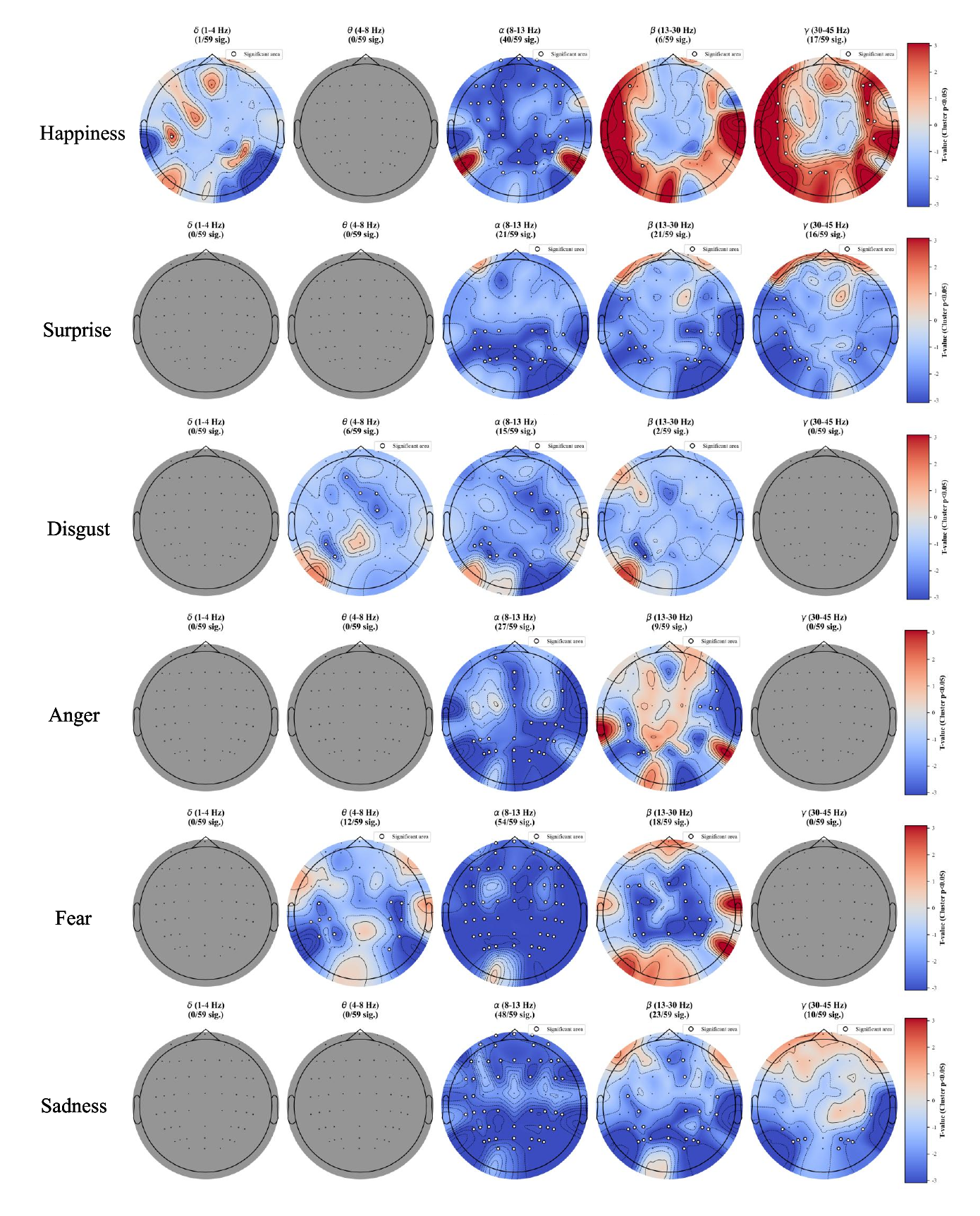}
	\caption{Event-related EEG PSD change T-maps (4-s window) for six emotions across five frequency bands. White circles indicate channels in significant clusters (cluster permutation test, $p<0.05$). Red/blue indicates PSD increase/decrease. The figure compares neural signatures across emotions.}
	\label{fig17}
\end{figure}

As shown in Fig.~\ref{fig17}, widespread $\alpha$ desynchronization emerged as a reliable common marker of generic cortical engagement across emotions, whereas other frequency bands exhibited highly emotion-specific modulations. To illustrate these diverse patterns, we highlight two representative emotions:

\begin{itemize}
    \item \textbf{Fear:} Elicited the most profound $\alpha$ suppression (54 electrodes) across almost the entire scalp, coupled with significant $\beta$ and $\theta$ desynchronization, indicating intense cortical arousal and heightened vigilance \cite{chien2017oscillatory,bacigalupo2022alpha}.
    \item \textbf{Happiness:} Characterized by moderate frontoparietal $\alpha$ desynchronization (40 electrodes), uniquely accompanied by prominent $\gamma$ synchronization (17 electrodes) and $\beta$ enhancement in temporo-frontal areas, reflecting positive affect processing. The consistency of this pattern with prior observations from SEED-VII further supports the neurophysiological validity of the proposed event-centered annotations~\cite{jiang2024seed,schwartz1997neuroanatomical,jatupaiboon2013real}.
\end{itemize}

To further validate our choice of a 4-s window, we replicated the analysis using adjacent 3 s and 5 s intervals. As shown in Supplementary Figs.~S1 and S2, altering the window duration substantially diluted key neural signatures. Notably, within the 5 s window ([$t_{event}$ - 2 s, $t_{event}$ + 3 s]), the localized patterns for Surprise and Disgust vanished entirely, while Fear-related $\alpha$ suppression and Happiness-related $\gamma$ enhancement were markedly attenuated. This confirms that the 4-s window optimally isolates the event-centered neural dynamics.

\begin{figure}[htbp]
	\centering
	\includegraphics[width=\columnwidth]{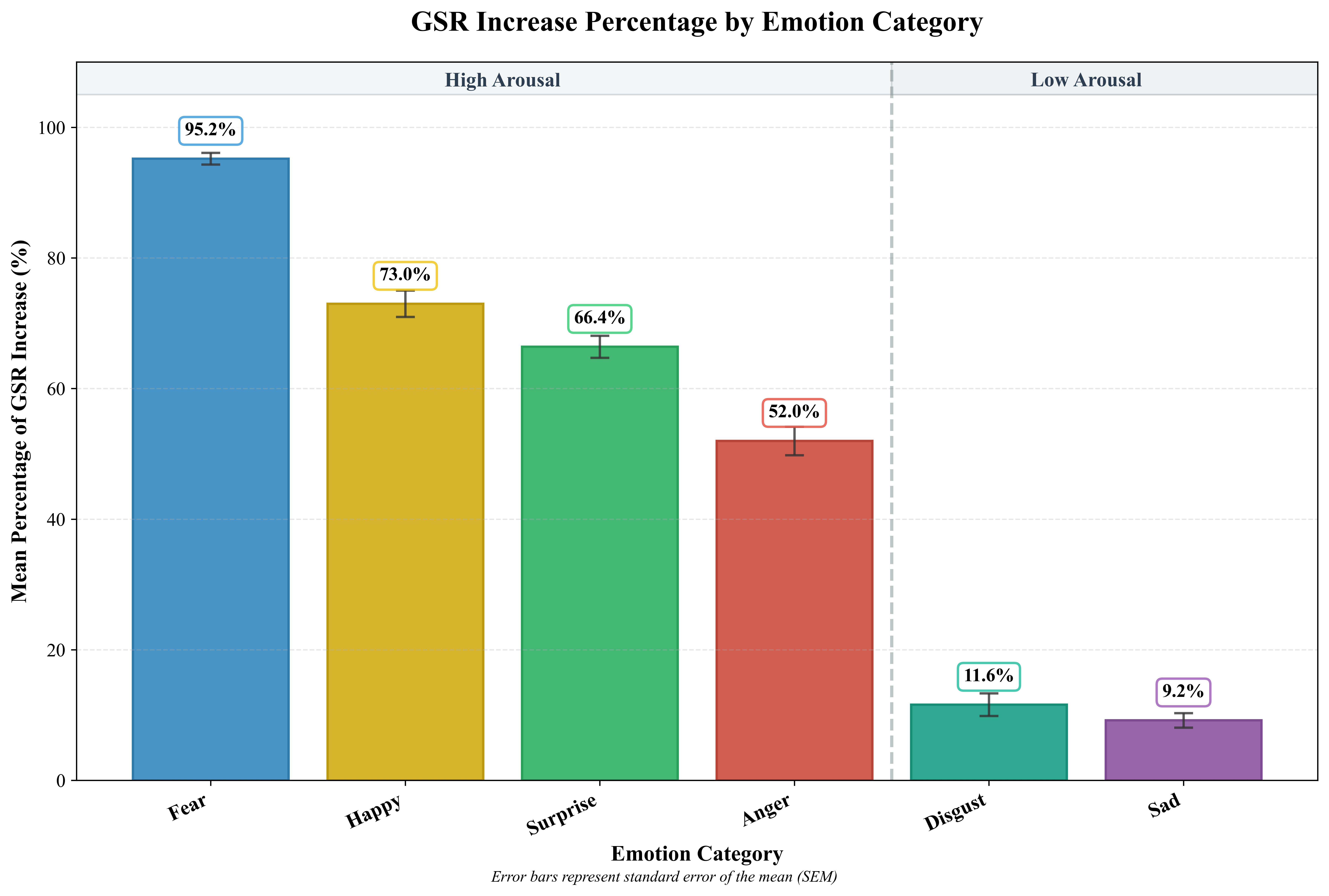}
	\caption{GSR increase percentage for each of the six basic emotions.}
	\label{fig:gsr_percentages}
\end{figure}

\textbf{Autonomic Nervous System (GSR):} Emotional arousal is tightly coupled with sympathetic nervous system (SNS) activity, which directly drives variations in Galvanic Skin Response (GSR) via sweat gland innervation. Transient, event-related surges in this continuous signal—known as Skin Conductance Responses (SCRs)—serve as robust objective biomarkers for high-arousal states. To validate the temporal precision of our annotations, we quantified the occurrence rate of significant SCRs. Specifically, a physiological surge was classified as a valid SCR if its amplitude exceeded the standard threshold of $0.05\mu\text{S}$ within a 2-second window following the annotated peak ($t_{event}$). As visualized in Fig. \ref{fig:gsr_percentages}, a sharp physiological dichotomy emerged between emotion categories. High-arousal emotions induced massive, immediate autonomic surges, with the mean valid SCR occurrence rate reaching an overwhelming 95.2\% for Fear, followed by Happiness (73.0\%), Surprise (66.4\%), and Anger (52.0\%). Conversely, low-arousal emotions (Disgust and Sadness) exhibited flat baselines (11.6\% and 9.2\%, respectively). This distinctive arousal pattern strongly corroborates that our immediate-recall paradigm successfully isolates genuine physiological peaks from temporal noise.

\section{Benchmark Experiments}

\subsection{Setup and Baselines}
To quantify the benefit of our annotation strategy on downstream machine learning tasks, we conducted emotion classification using a rigorous Leave-One-Subject-Out (LOSO) cross-validation protocol. We benchmarked \textbf{FIRMED} (trained on localized 4-s windows) directly against \textbf{FIRMED-Whole}, which partitions each complete trial into consecutive, non-overlapping 4-s segments, each inheriting the single trial-level emotion label---mirroring the standard segmentation protocol widely adopted by DEAP~\cite{koelstra2011deap}, SEED~\cite{zheng2015investigating}. We evaluated eight established classifiers ranging from traditional methods to advanced neural architectures.

\subsection{EEG-Based Emotion Recognition Results}
For EEG, we extracted Differential Entropy (DE) features across five frequency bands. As presented in Table \ref{tab:eeg_loso}, models trained on the finely annotated FIRMED dataset consistently outperformed those trained on FIRMED-Whole across all eight classifiers. The average classification accuracy improved by 3.8\% points (38.5\% vs. 34.7\%). The advanced transformer-based model MAET achieved the highest single-modality accuracy of 46.1\%.

Notably, the magnitude of improvement aligns strictly with the models' spatio-temporal modeling capacities. Traditional algorithms (e.g., SVM, k-NN) exhibit modest gains (+2.6\% to +2.8\%), whereas advanced deep architectures like MD$^2$GRL and MAET fully leverage the temporally purified supervision, achieving superior performance boosts of +4.6\% and +4.8\%, respectively. These uniform yet model-dependent gains suggest that mitigating temporal label noise improves feature discriminability.

\begin{table}[t] 
\centering
\caption{EEG emotion recognition performance under LOSO evaluation. Peak-centered annotations yield steady and model-dependent gains.}
\label{tab:eeg_loso}
\renewcommand{\arraystretch}{1.1} 
\begin{tabular}{lccc}
\toprule
Classifier & FIRMED (\%) & FIRMED-Whole (\%) & $\Delta$ \\
\midrule
SVM~\cite{wang2011eeg}       & 26.8 & 24.2 & {+2.6} \\
k-NN~\cite{li2018emotion}       & 30.3 & 27.5 & {+2.8} \\
MLP~\cite{riedmiller2014multi}        & 35.7 & 32.4 & {+3.3} \\
LSTM~\cite{graves2012long}       & 37.9 & 34.1 & {+3.8} \\
EEGNet~\cite{lawhern2018eegnet}     & 41.6 & 37.5 & {+4.1} \\
RGNN~\cite{zhong2020eeg}       & 43.5 & 39.2 & {+4.3} \\
MD$^2$GRL~\cite{tang2024multi}  & 45.2 & 40.6 & {+4.6} \\
MAET~\cite{jiang2024seed}       & \textbf{46.1} & \textbf{41.3} & \textbf{+4.8} \\
\midrule
Average    & 38.5 & 34.7 & \textbf{+3.8} \\
\bottomrule
\end{tabular}
\end{table}

\subsection{Multimodal Fusion and Window Ablation}
Using the best-performing MAET as the classifier backbone, we further 
evaluated multimodal fusion configurations on FIRMED, as summarized in 
Table~\ref{tab:multi-window}. Modality-specific features were extracted 
independently: GSR derivative skewness, ECG RMSSD, 
PPG amplitude change, and AU-based facial features as 
described in Section \ref{sec3.3}. For each fusion setting, the respective feature vectors were z-score normalized per subject and concatenated into a 
unified representation, which was then fed into MAET operating in its 
single-input mode. The training protocol and LOSO evaluation remained 
identical across all configurations.

All multimodal settings consistently outperformed the EEG-only baseline 
on FIRMED. Among individual peripheral modalities, GSR provided the 
largest single-channel boost (46.1\% $\to$ 51.3\%). ECG and PPG 
contributed moderate gains (48.9\% and 48.2\%, respectively). Facial AU 
features yielded 47.5\%, indicating that behavioral surface cues encode 
complementary affective information beyond physiological channels. 
Combining physiological peripherals (EEG+GSR+ECG+PPG) reached 53.4\%, 
and introducing Face into the GSR-augmented configuration 
(EEG+GSR+Face) achieved 52.1\%. The full five-modality fusion 
(EEG+GSR+ECG+PPG+Face) attained the overall peak accuracy of 
\textbf{54.8\%}. Crucially, the consistent advantage of FIRMED over 
FIRMED-Whole persisted across all fusion configurations (Table~\ref{tab:multi-window}), 
suggesting that the benefit of peak-centered annotation generalizes across different modality settings. Notably, under FIRMED-Whole, adding facial features slightly decreased accuracy (46.2\% $\to$ 45.7\%), likely because near-neutral expressions in non-salient segments inherit the trial-level label and introduce inter-modal conflict during fusion.
\begin{table}[t]
\centering
\caption{MAET classifier performance under early-fusion multimodal 
configurations and ablation on event-window size.}
\label{tab:multi-window}
\renewcommand{\arraystretch}{1.1} 
\resizebox{\columnwidth}{!}{
\begin{tabular}{lcc}
\toprule
Experimental Setting & FIRMED (\%) & FIRMED-Whole (\%) \\
\midrule
\textit{Modality Combinations} & & \\
EEG only              & 46.1 & 41.3 \\
EEG + PPG             & 48.2 & 43.1 \\
EEG + ECG             & 48.9 & 43.6 \\
EEG + Face            & 47.5 & 40.3 \\
EEG + GSR             & 51.3 & 44.8 \\
EEG + GSR + Face      & 52.1 & 45.3 \\
EEG + GSR + ECG + PPG & 53.4 & 46.2 \\
EEG + GSR + ECG + PPG + Face & \textbf{54.8} & 45.7 \\
\midrule
\textit{Window Length Ablation (EEG)} & & \\
3-s window ($t \pm 1.5s$) & 45.4 & 41.5 \\
\textbf{4-s window ($t \pm 2.0s$)} & \textbf{46.1} & \textbf{41.3} \\
5-s window ($t \pm 2.5s$) & 45.7 & 42.1 \\
6-s window ($t \pm 3.0s$) & 44.8 & 41.8 \\
\bottomrule
\end{tabular}
} 
\end{table}
\section{Access, Reproducibility, and Ethics}
We construct and make FIRMED available for academic research use. Benchmark protocols, metadata, annotations, preprocessing scripts, feature extraction code, and evaluation scripts are publicly available at \url{https://github.com/NIlab666/FIRMED}. Due to privacy concerns associated with physiological and facial recordings, raw data are available to qualified academic researchers through a controlled-access process under a standard data use agreement, which prohibits re-identification and redistribution. The study was approved by the institutional ethics committee, and all participants provided written informed consent for data collection and controlled sharing.

\section{Conclusion}
We introduced FIRMED, a peak-centered multimodal benchmark for dynamic video emotion recognition. By adopting an immediate-recall paradigm, FIRMED reduces temporal label noise and enables localized supervision aligned with objective neural and autonomic arousal states. Benchmark results across diverse classifiers and multimodal settings show clear gains over conventional whole-trial labeling. FIRMED provides a practical testbed for temporally localized emotion recognition, multimodal fusion, subject-independent affect modeling, and future studies on annotation reliability and label-noise-robust learning.

FIRMED also has several limitations. The current dataset remains moderate in size. The collected Big Five personality data were not utilized in the current analyses and will be explored in future work to model individual differences in emotional reactivity. In addition, the current benchmark relies on a fixed event window, which may not fully capture emotion-specific temporal dynamics. Future work will expand subject diversity and explore more adaptive windowing strategies.
\bibliographystyle{ACM-Reference-Format}
\bibliography{myrefs}

\end{document}